\documentclass{article}

\usepackage{arxiv}

\usepackage[utf8]{inputenc} % allow utf-8 input
\usepackage[T1]{fontenc}    % use 8-bit T1 fonts
\usepackage{hyperref}       % hyperlinks
\usepackage{url}            % simple URL typesetting
\usepackage{booktabs}       % professional-quality tables
\usepackage{amsfonts}       % blackboard math symbols
\usepackage{nicefrac}       % compact symbols for 1/2, etc.
\usepackage{microtype}      % microtypography
\usepackage{graphicx}
\usepackage{tabularx}
\usepackage{multirow}

\title{A Discussion and Comparative Study on Security and Privacy of Smart Meter Data}

\author{ \href{https://orcid.org/0000-0002-9319-3408}{\includegraphics[scale=0.06]{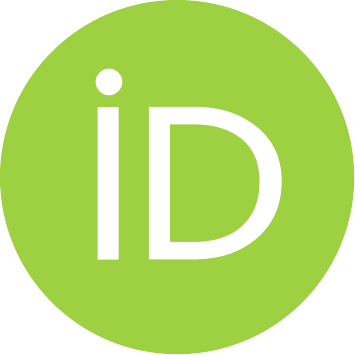}\hspace{1mm}Jatinder Kumar}\thanks{The authors would like to thank National Institute of Technology Kurukshetra, India for financially supporting this research work.} \\
	Department of Computer Applications\\
    National Institute of Technology\\
	Kurukshetra, India \\
	\texttt{jatinder\_61900097@nitkkr.ac.in} \\
	\And
	\href{https://orcid.org/0000-0002-8053-5050}{\includegraphics[scale=0.06]{orcid.pdf}\hspace{1mm}Ashutosh Kumar Singh} \\
		Department of Computer Applications\\
      National Institute of Technology\\
	Kurukshetra, India \\
	\texttt{ashutosh@nitkkr.ac.in} \\

}

\renewcommand{\shorttitle}{A Discussion and Comparative Study on Security and Privacy of Smart Meter Data}

\hypersetup{
pdftitle={A template for the arxiv style},
pdfsubject={q-bio.NC, q-bio.QM},
pdfauthor={David S.~Hippocampus, Elias D.~Striatum},
pdfkeywords={First keyword, Second keyword, More},
}

\begin{document}
\maketitle

\begin{abstract}

Cloud computing comes with a lot of advanced features along with privacy and security problem. 
Smart meter data takes the benefit of cloud computing in the smart grid. User's privacy can be compromised by analyzing the smart meter data generated by household electrical appliances. 
The user loses control over the data while data is shifted to the cloud.
This paper describes the issues under the privacy and security of smart meter data in the cloud environment. We also compare the existing approaches for preserving the privacy and security of smart meter data.

\end{abstract}
%https://www.overleaf.com/project/6129194805bd1181f52dde18

% keywords can be removed

\section{Introduction}
In recent years, cloud computing has been an emerging technology that is revolutionizing IT infrastructure and flexibility. Cloud computing refers to the delivery of on-demand computing resources over the network (typically the internet)\cite{rishabh2021arxiv}\cite{aman2020}. Resources can be processors, storage, software, network, and many more. It drastically transforms the means of computing. It allows users to leverage the computing resources for the required time on the “pay-as-you-go” model\cite{ishu2020ieee}-\cite{saxena2021isj}. Cloud computing also makes it possible to access resources from anywhere, while in the traditional computer system, you have to be there, where the physical resource is located \cite{ashutosh2021ipdc}-\cite{jkumar2021etri}. Cloud users process and store their data on publicly owned or outsourced data centers. The location of these data centers may be across the world. Companies, organizations, governments, or others are keeping their applications up and faster through this technology \cite{jkumar2021is}-\cite{jkumar2020cc}. It also helps the clients in avoiding up-front infrastructure costs. Many organizations are adopting cloud computing due to its characteristics like scalability, availability, robustness, pay-as-you-go, etc\cite{jkumar2020sc}-\cite{singh2019}.
Cloud computing can be classified into two types of models, i.e., deployment model and service model. Organizations choose when, how, and which model is appropriate for use based on their specific requirements.
\begin{itemize}
  \item \textbf{Deployment Model} \\
  There are four deployment models, and organizations choose one of them, depending on their specific requirements. These models are public, private, hybrid, and community cloud models. A third-party organization owns a public cloud and is accessible for public use or people under an organization. The private cloud offers a more controlled environment where access to IT resources is more centralized within the business. This model can be externally hosted or can be managed in-house. A hybrid cloud environment can be deployed for an organization seeking the benefits of both private and public cloud deployment models\cite{jkumar2019ieee}\cite{jkumar2018fgcs}. A cloud that is originated to serve a common function or purpose is called a community cloud. The deployment model is depicted in Fig. 1.
\begin{figure}[!htbp]
    \centering
    \includegraphics[width=6.2cm]{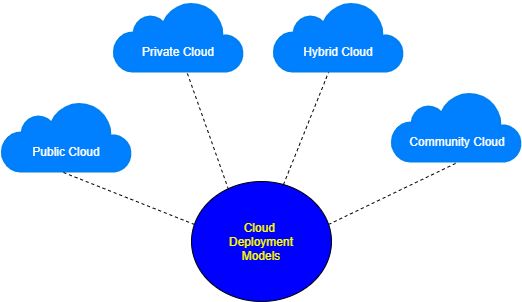}
    \caption{Deployment model of cloud computing}
   
\end{figure}  

  \item \textbf{Service Model} \\
  There are the main types of cloud service models- Infrastructure-as-a-service (IaaS), Software-as-a-service (SaaS), and Platform-as-a-service (Pa-as). The service model of cloud computing is depicted in Fig. 2. As their name, they offer services of Infrastructure, software, and platform, respectively. In IaaS, infrastructure components such as hardware, servers, storage, and other services are provided\cite{jkumar2016}\cite{saxena2021arxiv}. In SaaS, the services of any software are provided over the internet. Users can use this software without installing it on his machine. In PaaS, cloud computing providers deploy the infrastructure and software framework, but businesses can develop and run their applications. Web applications can be created quickly and easily via PaaS.
  \begin{figure}[!htbp]
    \centering
    \includegraphics[width=6cm]{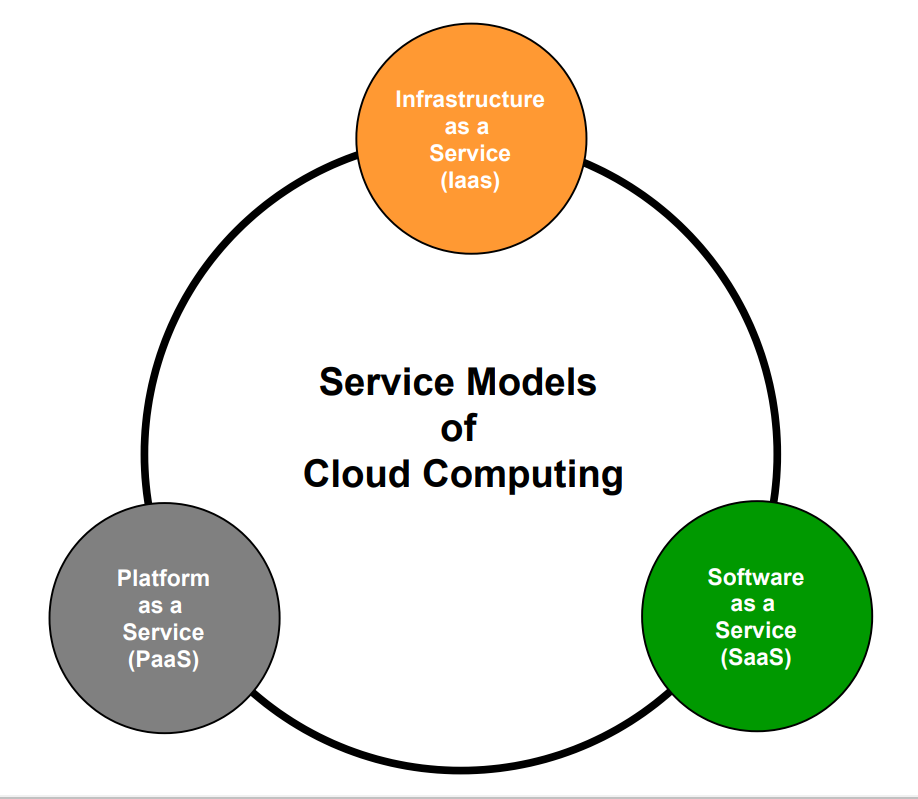}
    \caption{Service model of cloud computing}
\end{figure} 
\end{itemize}
\section{Motivation}
\label{sec:headings}
In 2003, over 10 million people in North America got affected due to electrical blackout. Blackout happens due to load imbalance and lack of effective real-time analysis \cite{lu2012}. Electricity load should be matched with the electricity generation of the power grid. The problem of real-time analysis of every individual's load and is done by introducing a smart grid. According to Electric Power Research Institute (EPRI), a fully developed smart grid can save up to \$2 trillion in 20 years where the deployment cost between \$338 to \$476 billion \cite{jindal2020}. By the end of 2018, the U.S. had over 86 million smart meters installed. In 2017, there were 665 million smart meters installed globally. Revenue generation is expected to grow from \$12.8 billion in 2017 to \$20 billion by 2022 \cite{market2020}. Smart grids have been developed by integrating the traditional power grid to efficiently monitor power generation and energy usage through two-way communications between the consumer and utility provider. By smart grid technology, consumers can be notified for energy pricing of real-time data analysis. The utility also processes and analyzes the data for the generation and distribution of energy in the regions. Smart grid can be effectively used for essential tasks such as accounting, theft detection, and optimization.
Utility loses \$58.7 billion per year worldwide \cite{newswire2014}.

National Institute of Standards and Technology (NIST) presents the smart grid model consists of seven domains: Transmission, Distribution, Operations, Generations, Market, Consumer, and Service provider \cite{abdallah2016} . A smart meter is an important component of the smart grid that collects real-time data and transmits it to the utility provider. Smart meters are located and divided by different localities. Each locality consists one base station which aggregates data of all smart meter and sends to control center. By analyzing smart meter data, future loads can be predicted by deep learning algorithms, and energy can be saved on the consumer side and reduce energy at the production side \cite{wang2018}. Machine learning and deep learning can be employed in smart grids to perform load prediction and energy patterns \cite{hosein2017} \cite{tanwar2019}. An effort must be made to overcome the problem of ‘secure transmission and storing of smart meter data on the outsourced cloud with privacy-preserving.

\section{Related Work}
Data generated by smart meter is of huge amount as it sends reading at after every minute. This data is terabyte-level big data. So, storing and analyzing data is a very difficult task. 
Many encryption algorithms are proposed to hide the sensitive data of smart meter users \cite{zhang2017}-\cite{okay2016}.
Z. Zhang et al. \cite{zhang2017} propose privacy-friendly cloud storage (PCS) scheme for storing smart meter (SM) data on an outsourced cloud. In this scheme, the author uses two clouds to preserve consumers' privacy and homomorphic encryption to encrypt the smart meter's data. After every minute, the smart meter sends the data to the cloud in encrypted form with asymmetric cryptography. Both clouds decrypt the data and encrypt it again, with homomorphic encryption having the property of performing operations on encrypted data. After encryption, all the data is stored on one cloud, which is then used for analysis. Stored data is of huge amount. Hadoop is used for handling big data generated by the smart meters of a large number of houses.

O. Rafik and S. Senouci \cite{merad2020} propose an Efficient and Secure Multidimensional data Aggregation (ESMA) scheme, which consists of an FN which works as a data aggregation node. A FN uses paillier homomorphic encryption to aggregate the smart meter’s multidimensional data and convert it in to a single ciphertext. ESMA also handles fault-tolerant. If a SM fails to send its reading, then the final result will not be affected. ESMA verifies the authenticity of data sent by the SM and FN to the control center. The query is also handled by the Control Center (CC) within the summation operation. Authors evaluate the ESMA scheme on 500 SM and 40 types of data and get affordable compared to previous schemes. The main disadvantage of this scheme is that this scheme can’t handle the data of a large number of houses because that would be of Tera-byte level. The summary of related works is shown in Table 1.

\begin{table}[!htbp]
    \centering
		\caption{Summary of Security and Privacy Preserving Review}

		\small\addtolength{\tabcolsep}{-4pt}
			\begin{tabular}{|l|l|l|l|}\hline 
				\textbf{\textit{Literature }} & 
				\textbf{\textit{Approach}}&
					\textbf{\textit{Pros}} &  
						\textbf{\textit{Cons}} \\
						\textit{\textbf{Reference}} & & & \\
						\hline  \hline

			Z. Zhang et al. \cite{zhang2017}  & • Paillier  HE 
 & • Privacy-preserving & • High computation
 \\ 
  & • Hadoop for query & • Secure Statistics over  & \& communication cost \\
& processing & encrypted data & \\
 \hline
A. Abdallah and   & • Lattice based HE & • Data aggregation without & • Communication cost 
		\\
		X. Shen \cite{abdallah2016} & based data aggegation & involving SM and BS & is high in HAN \\
		\hline
T. Wang et al. \cite{wang2018}& • Fog computing is  & •  Whole data can’t be  & • Doesn’t support for  \\ & introduced between & accessed if data at one & large scale data \\
 & • Data partition using Hash- & • Ownership and management & • Computation cost is \\
 & Solomon code algorithm & of data at cloud can be done & high \\ \hline

R. Lu et al. \cite{lu2012}  & • Multidimensional data   & • Maintain privacy with   & • Storage and analysis  \\ 
& aggregation converted into  & authentication of every  reading & problem at OC for large  \\
& single ciphertext &  &  HANs\\ \hline

	C. Fan et al. \cite{fan2013} &  •	Diffie-Hellman key  & •	Secure batch verification  & •	Time cost is high  \\ &  exchange algorithm & system & \\ \hline

M. Badra \& S.  &  •	Data aggregation with   &  •	Saves from unauthorized  & •	Computation cost is    \\ Zeadally \cite{badra2017} & lightweight symmetric HE & modification of the data & high \\ 
& • ECDH key exchange & & •	Latency is high\\
& algorithm  &  &  \\
\hline

O. Boudia et  & •	Aggregator node works  & •	Verification of data  & •	Storage is insufficient  \\ 
 al. \cite{boudia2017} & for data aggregation  & at Aggregator node & for storing smart meter    \\ 
 & of different HANs & •	Multidimensional data & data at control center \\
 & & is aggregated  & \\
 \hline

S. Sarkar et  & •	Introduce fog node  & •	Improved latency of  & •	More hardware cost by  \\ al. \cite{sarkar2015}&  between mobile terminal  & applications &  adding fog nodes with   \\
 &  nodes and cloud data  & •	Less power consumption & mobile terminal nodes \\
 & centers & of resources & \\
 \hline

			\end{tabular}

	\end{table}

\begin{table}[!htbp]
    \centering

		\small\addtolength{\tabcolsep}{-4pt}
			\begin{tabular}{|l|l|l|l|}\hline

O. Rafik and S. & •	Some smart meter are  & •	Multidimensional data is  & •	Analyzing large data at  \\ Senouci \cite{merad2020} & connected with one FN and  & aggregated with  &  control center takes more   \\
 & every FN is connected to  & fault-tolerant. & time \\ 
 & Paillier HE & •	Batch verification & \\
 &  & and query response & \\
 \hline
 
X. Zuo et al. \cite{zuo2020}& •	ElGamal cryptography  & •	Resist from attacks at  & •	Latency is high \\ & without trusted  & Gateway and control  &  •	Gateway does not  large   \\
 & Authority & center & number of SM is connected \\
 \hline
F. Okay and S. & •	Fog computing is  & •	Increased privacy and  & •	Doesn’t support  \\  Ozdemir \cite{okay2016} & used in smart grid. & improved latency &  multidimensional data  \\ \hline
 
			\end{tabular}

	\end{table}

\vspace{1cm}

\section{Research Gaps}
On the basis of the literature review, the following research gaps are identified.
\begin{enumerate}
  
  \item More computational and communicational cost.
  \item Less privacy preservation of data.
  \item Analysis of data of millions of customer with traditional tools is difficult.
  \item Querying of data on the cloud is not secured.
\end{enumerate}

\section{Research Objectives}
To fill the identified study gaps, the goal described below is developed.
\begin{enumerate}
  \item To reduce communicational and computational cost.
  \item To improve privacy using encryption algorithms.

  \item To improve latency of transmission of data.
  \item Secure query processing on the cloud.

\end{enumerate}

%\vspace{2cm}

\end{document}